\documentclass[
aps,
pra,
twocolumn,
superscriptaddress,
amsmath,
amssymb,
showpacs
]{revtex4-2}

\usepackage[utf8x]{inputenc}
\usepackage{color}
\usepackage{amsmath}
\usepackage{amssymb}
\usepackage{amsfonts}
\usepackage{amsthm}
\usepackage{float}
\usepackage{bm}
\usepackage{latexsym}
\usepackage{hyperref}
\usepackage{multirow}
\usepackage[capitalize]{cleveref}
\usepackage{graphicx}
\hypersetup{
	colorlinks,
	citecolor=blue,    filecolor=blue,
	linkcolor=blue,     urlcolor=blue
}

\begin{document}

\title{Optical modelling of shaped laser pulses in plasma} 

\author{Igor A. Andriyash}
\email[]{igor.andriyash@ensta-paris.fr}
\affiliation{Laboratoire d’Optique Appliquée (LOA), CNRS, École polytechnique, ENSTA, Institut Polytechnique de Paris, Palaiseau, France}

\author{Cedric Thaury}
\affiliation{Laboratoire d’Optique Appliquée (LOA), CNRS, École polytechnique, ENSTA, Institut Polytechnique de Paris, Palaiseau, France}

\date{\today}

\begin{abstract}

This work provides a brief review of the numerical methods for modelling the optical propagation of an ultrashort, intense laser pulse in plasmas and ionizable gases. These methods are implemented in an open-source simulation toolkit Axiprop, which is now actively used for design studies in the context of laser plasma acceleration of electrons (LPA). We present two examples of such studies: optical plasma waveguide generation, and phase-locked flying-focus LPA. These tools enable the identification of complex effects impacting both energy deposition during waveguide formation, and the pulse propagation dynamics that governs wakefield structure, ultimately determining the properties of the accelerated electron bunches. The developed tools and presented simulation designs are relevant to experiments on laser plasma interactions at modern high power laser facilities.

\end{abstract}

\maketitle
\normalfont

\section{Introduction}

Development and growing accessibility of high-power, ultra-fast lasers have established the research field of relativistic laser plasmas. Today, such lasers produce plasma sources of relativistic particles and x-ray/$\gamma$-ray beams, and pave the way towards such applications as free-electron lasers, colliders, and observations of strong-field QED phenomena. Harnessing such laser powers requires advanced optical tools to shape and guide these beams. Laser pulses can be shaped in many ways, for example, using an axicon, axilens or axiparabola \cite{Smartsev:2019} one can produce quasi-Bessel beams with long focal lines and superluminal on-axis group velocity \cite{Oubrerie:JOPT2022}. Alternatively, by introducing the pulse front curvature and spectral chirp, one can control the pulse velocity \cite{Sainte-Marie:Optica2017}, which makes such techniques extremely promising for the phase-locked laser plasma acceleration \cite{Caizergues:NatPhot2020,palastro2020dephasingless}. Moreover, the pulses with long focal line are often considered for optical shaping of gas targets. For example, using a few-mJ quasi-Bessel pulse one can produce a plasma waveguide for the main LPA driver via the so-called HOFI channel formation \cite{Shalloo:PRE2018,Oubrerie:LSA2022,Mewes:PRR2023,Picksley:PRL2024}.

Numerical studies of the relevant laser-plasma conditions are commonly carried out with particle-in-cell (PIC) modelling \cite{BirdsallLangdon}. However, in the cases of large shaped beams and long centimetric propagation, the computation cost of standard PIC simulations becomes prohibitive, and in the past two decades a number of remarkable numerical techniques have been developed to address this challenge. Today one can accelerate such calculations by considering the quasi-static approximation with an enveloped laser field \cite{Sprangle:PRL1990,Mora:PoP1997}, decomposing the full 3D geometry into a few cylindrical ones \cite{Lifschitz:JCP2009}, introducing Lorentz boost \cite{Vay:PRL2007}, and combining these techniques \cite{Lehe:PRE2016,Li:CPC2021,Massimo:PPCF2025}.

When designing a simulation with high-power quasi-Bessel beams one operates on two very different scales. The high intensity central part of the beam generates and drives the non-linear plasma typically localized over a few tens of micrometers around the beam axis. In cases of highly non-linear relativistic plasmas (e.g. LPA), this central part requires kinetic plasma simulations, that can also be coupled to the large-aperture analytical or numerical optical propagators, e.g. via total-field/scattered-field (TFSF) techniques \cite{Schneider:IEEETAP2004,Weichman:CPC2024}, or by considering the gridless plasma descriptions such as QSA-PIC \cite{Baxevanis:PRAB2018,FerranPousa:2019}. On the other hand, the major part of the quasi-Bessel beam energy is carried by the low-intensity converging-diverging fields that have millimetric apertures, and interact with plasma as the optical medium. Moreover, since the on-axis field is is continuously formed by the by converging wavefront, its group velocity is mainly defined by this optical propagation and may be considered using faster scalar field propagation simulations with plasma presented as a non-linear medium \cite{goodman2005introduction,couairon2011practitioner}. Such methods are well developed, and can integrate various optical properties of the media \cite{couairon2011practitioner}, and can be coupled with modern optimization algorithms \cite{Miller:ML2026}. Today optical modelling has become an important part of the LPA numerical studies, being essential in the experimental design \cite{Monzac:PRR2024,Oubrerie:LSA2022}.

In this work, we present numerical tools for optical modelling of short intense laser pulses in field-ionised weakly-relativistic plasmas. These tools are based on the standard optical propagators that can be coupled to models of plasma generation and non-linear currents. In Sec.~\ref{sec:basic}, we review concepts and methods for optical propagation computations, starting from basic terms, main equations, plasma models, and implementation techniques. In Sec.~\ref{sec:sims}, we demonstrate the use of the developed tools for simulation design studies of HOFI channel generation for guided LPA (Sec.~\ref{subsec:channel}) and phase-locked LPA (Sec.~\ref{subsec:lpa}), both harnessing the axiparabola mirror concept.

\section{Basics of optical calculations}\label{sec:basic}

The evolution of the electromagnetic field from one point in time and space to another is described by \textit{Maxwell's equations} for the electric and magnetic fields $ \mathbf{E}$ and $ \mathbf{H}$:
\begin{subequations}
 \begin{align}
  & \nabla \times \mathbf{E} = - \partial_t \mathbf{B}\,,\label{Maxwell1}\\
  & \nabla \times \mathbf{H} = \partial_t \mathbf{D} + \mathbf{J}\,,\label{Maxwell2}\\
  & \nabla \mathbf{D} = \rho\,,\label{Maxwell3}\\
 & \nabla  \mathbf{B} =0\,.\label{Maxwell4}\\
 & \mathbf{D}  = \epsilon_0 \mathbf{E} + \mathbf{P} \,, \label{Maxwell5}\\
 & \mathbf{H}  = \cfrac{1}{\mu_0} \mathbf{B} - \mathbf{M} \,,\label{Maxwell6}
 \end{align}
\end{subequations}
where $\epsilon_0$ and $\mu_0$ are the electric permittivity and magnetic permeability of vacuum. The contribution of the medium is defined by the charge and current densities of the free charge carriers $\rho$ and $\mathbf{J}$, as well as by the resulting displacement $\mathbf{D}$ and magnetic induction $\mathbf{B}$. In calculations with bounded charges, the medium is described by its polarization $\mathbf{P}$ and magnetization $\mathbf{M}$, which are commonly expressed as series of the electric and magnetic field components, with coefficients defining the electric and magnetic susceptibilities. In medium with free charges such as plasmas, the response arises from electron oscillations presented by terms $\rho$ and $\mathbf{J}$. For our model we discard the macroscopic terms $\mathbf{P}$ and $\mathbf{M}$ in \cref{Maxwell5,Maxwell6}, and develop plasma description based on calculations of current densities.


Let us assume propagation along the positive direction of the $z$-axis, and adopt a \textit{time-domain} representation, considering the field to be a function of time in the transverse $XY$ plane, at a given $z$ position. Combining \cref{Maxwell1,Maxwell2,Maxwell3,Maxwell4,Maxwell5,Maxwell6}, and assuming the free charges condition we can write the propagation equation for the electric field:
\begin{align}
	& \partial_z^2 \mathbf{E}  = - \nabla_\perp^2 \mathbf{E} + \frac{1}{c^2} \; \partial_t^2 \mathbf{E} + \mu_0\; \partial_t\mathbf{J}  + \frac{1}{\epsilon_0}\nabla \rho\,,\label{PropE12}
\end{align}
where the transverse Laplace operator reads ${\nabla_\perp^2\equiv \partial_x^2 + \partial_y^2}$ in Cartesian coordinates, and ${\nabla_\perp^2\equiv r^{-1} \partial_r r \partial_r  + r^{-2} \partial_\theta^2}$ in cylindrical geometry.

A widely used approach in optics is to represent the field through its decomposition into a series of monochromatic waves, ${\mathbf{E}(t) = \sum_j \mathbf{E}_{\omega_j} \exp(-\mathrm{i}\omega_j t)}$, i.e. the temporal discrete Fourier transform (DFT). This decomposition allows for the linearization of the time derivatives in \cref{PropE12}. The resulting equation in spatial coordinates, can be further linearized, by considering the \textit{spatial-spectral decomposition} in the transverse plane. In Cartesian coordinates, this takes form of the standard two dimensional Fourier transform,
\begin{subequations}
\begin{align}
	\mathbf{E} (x, y) = \sum_{k_x}\sum_{k_y} \, \mathrm{e}^{i (k_x x + k_y y)}  \, \mathbf{E}_\omega \,,\label{def_fft2}
\end{align}
and in the cylindrical case it becomes the Fourier-Bessel transform,
\begin{align}
\mathbf{E}_\omega(r, \theta) =  \sum_m \; \mathrm{e}^{-i m \theta} \sum_{k_r}\; J_{m}(k_r r) \; \mathbf{E}_\omega(k_r, m) \,.\label{def_hdt}
\end{align}
\end{subequations}

Transforms \cref{def_fft2,def_hdt} linearize the transverse Laplace operator, $\nabla_\perp^2 \to -k_r^2$, with $k_r\equiv \sqrt{k_x^2 + k_y^2}$ for \cref{def_fft2}, and provides an ordinary differential equation for the spectral components:
\begin{align}
	& \partial_z^2 \hat{\mathbf{E}}  = - k_z^2 \hat{\mathbf{E}} - i \mu_0\; \omega \hat{\mathbf{J}} + \frac{1}{\epsilon_0}\widehat{\nabla \rho}\,,\label{PropE14}
\end{align}
where we denote $k_z = \sqrt{\omega^2/c^2 - k_r^2 }$, and $\hat{\mathbf{E}}$ stands for the full spatio-temporal spectral transforms, $\mathbf{E}(t, x,y,z) \to \hat{\mathbf{E}}(\omega, k_x, k_y, z)$ and $\mathbf{E}(t, r,\theta,z) \to \hat{\mathbf{E}}(\omega, k_r, m, z)$, for the Cartesian and cylindrical geometries respectively. 

Before we consider solutions of \cref{PropE14}, let us recall one more concept useful for the optical calculations, the \textit{field envelope representation}. For optical pulses with multiple field cycles, the proper temporal resolution of the oscillations typically requires a large number of sampling points. On the other hand, for a pulse with many cycles, $N_{cycles} \gg 1 $, the spectral signal is contained in a narrow bandwidth, $\delta\omega/\omega_0 \sim 1/N_{cycles}$, around the central frequency, $\omega_0=2\pi c/\lambda_0$, and, therefore, can be described with far fewer sampling points than its temporal counterpart. Practically, this can allows us to separate the oscillations from the waveform and work with the pulse envelope $\tilde{\mathbf{E}}$ defined as:
\begin{align}
  \mathbf{E}(\mathbf{r},t) = \textrm{Re}\left [ \tilde{\mathbf{E}}(\mathbf{r}, t) \exp(-i \omega_0 t)  \right]\,.  \label{def_envel}
\end{align}

Substituting \cref{def_envel} into \cref{PropE14}, is equivalent to replacing the high-frequency base $\omega$ with $\tilde{\omega}+\omega_0$, where $\tilde{\omega}$ corresponds to the spectral components of the envelope, which now contains both positive and negative frequencies. Typically, the envelope varies over timescales much larger than the field period and may be sampled with a moderate number of points. For unidirectional propagation, the actual field contains only positive frequencies, which limits the envelope sampling to $\Delta t \ge \pi/\omega_0$. Finer temporal sampling can still be considered to resolve the higher frequencies (e.g. harmonics), and in this case the arising negative frequencies should be discarded explicitly.

\subsection{Propagation in vacuum}\label{subsec:vacuum}

Let us first consider propagation of electromagnetic waves in free space. For this, we discard the medium response in \cref{PropE14}, which leads to the harmonic oscillator equation:
\begin{align}
	&  \partial_z^2 \hat{\mathbf{E}}  = - k_z^2 \hat{\mathbf{E}} \,,\label{PropEq_vacc1}
\end{align}
with the general solution,
\begin{align}
	& \hat{\mathbf{E}}(z)  = C_+ \mathrm{e}^{i k_z z } + C_- \mathrm{e}^{-i k_z z} \,, \label{PropEq_vacc2}
\end{align}
where the two terms define the waves propagating in the positive and negative directions along the $z$-axis respectively. For the waves travelling in the positive direction, the solution reads:
\begin{align}
	& \hat{\mathbf{E}}(z_1)  =  \hat{\mathbf{E}}(z_0)  \mathrm{e}^{i k_z (z_1-z_0)} \,. \label{PropEq_vacc3}
\end{align}
This result can be readily used to describe optical propagation over a distance $\Delta z=z_1-z_0$ by multiplying the field in the spectral domain by the \textit{phasor} factor,  $\exp (i k_z \Delta z )$. Such an approach can be used both for small propagation steps, and for long propagation distances, e.g. to propagate the field from the focussing optics to the focal position. 

The initial and final fields in \cref{PropEq_vacc3} intrinsically assume the same spatial resolution, thus requiring a high number of unnecessary spatial sampling points for the unfocussed field. This can be addressed in the limit, ${k_r\ll \omega/c}$, where one can linearize,
$$k_z=\sqrt{\omega^2/c^2 - k_r^2 }\simeq \omega / c - k_r^2  c  / 2 \omega\,,$$
which allows the oscillating and diffracting parts in \cref{PropEq_vacc3} to be separated. Application of the Fourier convolution theorem, then, yields the solution in spatial coordinates, well-known as the \textit{Fresnel diffraction integral}:
\begin{subequations}
\begin{align}
	 \mathbf{E}_\omega &(x, y, z_1)   =  \frac{\omega}{2 \pi i c  \Delta z}  \exp \left[ \frac{i \omega  \Delta z} {c}  \left (1+ \frac{ r^2 }{2\Delta z^2}\right) \right] \nonumber \\
	& \times  \int_{-\infty}^\infty  \int_{-\infty}^\infty   \mathrm{d}x' \; \mathrm{d}y' \;   \mathbf{E}_\omega (x', y', z_0) \;  \exp\left ( \frac{i \,\omega\, r'^2}{2 c\,\Delta z}   \right)	\nonumber  \\
	&  \times   \, \exp\left ( - \frac{i \omega}{c\Delta z}  \left(x x'+y y'\right) \right) \,.   \label{PropEq_vacc4}
\end{align}
where $r=\sqrt{x^2+y^2}$ and $r'=\sqrt{x'^2+y'^2}$. Transforming \cref{PropEq_vacc4} into cylindrical coordinates, and decomposing the field into angular Fourier modes, ${\mathbf{E}(\theta) = \sum_{m=-\infty}^\infty  \mathbf{E}_m \exp(-i m \theta)}$, we can write the expression for each mode as:
\begin{align}
	&\mathbf{E}_{\omega, m} (r, z_1)   =  \frac{\omega}{ i^{m+1} c  \Delta z}  \exp \left[ \frac{i \omega  \Delta z} {c}  \! \left (1 \! + \! \frac{ r^2 }{2\Delta z^2}\right) \right] \label{PropEq_vacc5}  \\
	& \times  \int_{0}^\infty  r' \mathrm{d}r'  \;   \mathbf{E}_{\omega,m} (r', z_0) \;  \exp \left ( \frac{i \,\omega\, r'^2}{2 c\,\Delta z}   \right)  \, J_m \left (  \frac{\omega r r'}{c  \Delta z}  \right )  \,. \nonumber
\end{align}\label{PropEq_vacc45}
\end{subequations}

The integrals on the right hand side of Eqs.~(\ref{PropEq_vacc45}) are the Fourier and Hankel transforms of the field multiplied by the diffracting phasor, $\exp\left( {i \,\omega\, r'^2}/{2 c\Delta z}\right)$. The spectral coordinates in Eqs.~(\ref{PropEq_vacc45}) are related to the spatial ones as, $k_x = \omega x / c \Delta z$, $k_y = \omega y / c \Delta z$, and $k_r = \omega r / c \Delta z$, which means that the resulting spatial samplings differ for each wavelength. Consequently, the reconstruction of the total field requires interpolation.

\subsection{Propagation in plasma}\label{subsec:plasma}

In plasma, the oscillatory motion of free electrons generates currents and charge densities, as described in \cref{PropE12}. For non-relativistic field intensities and short pulse durations, the ion motion and the magnetic force on electrons can be neglected, so the electron motion follows the electric field as, ${\partial_t v_e = -e E/m_e}$, and generates the current density ${J = -e n_{pe} v_e }$. Hence, the corresponding term in \cref{PropE12} becomes, ${\mu_0\; \partial_t J = \omega_{pe}^2/c^2 E}$, where the $n_{pe}$ is plasma electron density, and ${\omega_{pe} = \sqrt{e^2 n_{pe}/\epsilon_0 m_e}}$ is the cold plasma frequency. It follows from \cref{PropE14}, that for constant $n_{pe}$ this term alone modifies the phasor of \cref{PropEq_vacc3} as, $\exp (i \Delta z \sqrt{k_z^2 - \omega_{pe}^2/c^2} )$.

The second source term in \cref{PropE12}, $ \nabla \rho  / \epsilon_0$, represents the slower plasma response via charge density modulations, and in optical calculations it is usually discarded. By considering the charge continuity and rewriting the sources in \cref{PropE12} in the integral form, we can express both through the current density:
\begin{align}\label{SourcesEq1}
	& \mu_0 \; \partial_t\mathbf{J}  + \frac{1}{\epsilon_0}\nabla \rho =  \int_{-\infty}^{t} \frac{dt'}{\epsilon_0 } \left(\frac{1}{c^2} \; \partial_t^2 \mathbf{J} -   \nabla (\nabla \mathbf{J}) \right)\,.
\end{align}

Along the field polarization direction (e.g. the $x$-axis) the terms ${c^{-2}\partial_t^2 J_x}$ and ${\partial_x^2 J_x}$ in \cref{SourcesEq1} scale as ${ \sim J_x / \lambda_0^2 }$, and ${\sim J_x / R_0^2}$ respectively, where $ \lambda_0$ and $R_0$ are the wavelength and transverse size of the field. In the limit, $R_0\gg \lambda_0$, the second term vanishes. This approximation remains valid except for tightly focused, diffraction-limited cases, which require more complete vector models \cite{DebyeP:ADF1909,hu2020efficient}.

For simplicity, in the following discussion we consider a scalar field corresponding to the component along the direction of laser polarization, that we assume to be linear. With this we obtain the spectral domain propagation equation:
\begin{align}
	& \partial_z^2 \hat{E}  = - k_z^2 \hat{E} - i \mu_0\; \omega \hat{J} \,.\label{PropEq_envelopEq0}
\end{align}
Note that for pulses with more complex polarization, the field components are independent in vacuum, but may couple through the medium response, and in such cases the propagation equations should also be treated as coupled.

In optically transparent plasma, $\omega_{pe}\ll\omega_0$, the deviation of the field from the vacuum solution develops slowly, over scales much longer than the field wavelength. For the waveform that satisfies \cref{PropEq_vacc3}, ${\hat{E}(z)  =  \bar{E}(z)  \mathrm{e}^{i k_z z}}$, this means $| \partial_z \bar{E} | \ll k_z  |\bar{E}|$, which is known as the \textit{slowly evolving wave approximation} (SEWA). Substituting this waveform into \cref{PropEq_envelopEq0}, SEWA allows one to neglect the term, $\partial_z^2 \bar{E}$, yielding the first order differential equation,
\begin{align}
	& \partial_z \bar{E} = - \cfrac{\mu_0 \;\omega}{2k_z}\;  \bar{J} \,.\label{PropEq_envelopEq2}
\end{align}

In the general case of ionizable media and moderately relativistic electron motion, the relation between the field and the current is non-linear. In order to account for that, one has to integrate the equation of motion for plasma electrons as, ${p_e = e A + p_{e0}}$, where $A$ is the vector potential, and $p_{e0}$ is the particle initial momentum. Assuming purely electromagnetic fields, vector potential can be determined directly from the field as, $E = -\partial_t A $, and the current density becomes,
\begin{align} \label{current_motion_rel}
  J=-en_{pe} \cfrac{p_e}{m_e} \; \left(1 + \cfrac{p_e^2}{m_e^2 c^2}\right)^{-1/2}\,.
\end{align}

While pre-ionized electrons are at rest before the field arrival and follow the field as, ${p_e = e A}$, the electrons created via ionization appear in the field at rest at the moment of ionization $t_i$, and their subsequent motion follows, ${p_e = e (A- A (t_i)) }$. As laser pulse passes and its field vanishes, these photoelectrons retain the momenta ${p_{e0}= -eA (t_i)}$, thus, gaining kinetic energies, which is known as the optical field ionization (OFI) heating \cite{Burnett:JOSAB1989}. Depending on the ionization potential and laser ponderomotive potential, either multi-photon or tunnelling ionization takes place, and in the case of intense femtosecond infrared lasers, the tunnelling process typically dominates \cite{Keldysh:JETP1965,Popov:UFN2004}. In this work we employ the Ammosov-Delone-Krainov (ADK) model \cite{ADK:JETP} as formulated for PIC methods in \cite{ChenM:JCP2013}. For the sake of brevity we will omit the probability expressions for the DC and AC models, which can be found in Eqs.~(2) and (5) of \cite{ChenM:JCP2013}.

In calculations using field envelope, ionization is described using the AC probability rate. In this approach, instead of describing electrons with single-particle trajectories, we consider the averaged motion of electron ensembles generated over times comparable to the optical cycle. The integration constant, $p_{e0}$, should be considered as averaged over these ensembles; in practice this results in a small non-oscillating term, which does not couple with high frequency components. Note, that the associated electron heating can still be estimated from statistical considerations \cite{Schroeder:PhysRevSTAB2014,Tomassini:PoP2017}. Neglecting this residual drift we can calculate the enveloped current density as,
\begin{align} \label{current_motion_rel_env}
  \tilde{J}=-\cfrac{e^2 n_{pe} \tilde{A}}{m_e \sqrt{1 + |\tilde{A}|^2/ 2 m_e^2 c^2}} \,.
\end{align}

The loss of laser energy due to ionization can be accounted for by adding to \cref{current_motion_rel_env,current_motion_rel} the "ionization" current \cite{Rae:PRA1992,Nuter:PoP2011}:
\begin{align} \label{ioniz_current}
 J_{ioniz} = \cfrac{\Delta t}{|E|} \; \sum_Z \Delta n_{pe}^{(Z)}\; U_{ion}^{(Z)} \,,
\end{align}
where $U_{ion}^{(Z)}$ is the ionization potential of a given ion species with the charge $Z$, and $\Delta n_{pe}^{(Z)}$ is the electron density, produced by the field $E$ over the time step $\Delta t$.

\subsection{Implementation}\label{subsec:implement}

The calculation methods discussed in the previous sections are implemented as an open-source library named \textit{Axiprop} \cite{axiprop:2020}. The software is developed in Python, and it performs calculations on CPU and GPU processors through a number of available backends implemented using NumPy \cite{NumPy:Nature2020}, CuPy \cite{cupy:2017}, PyOpenCL \cite{PyOpenCL:2012} and Reikna \cite{reikna:manual} libraries. Axiprop is integrated as one of the engines for vacuum propagation in the collaborative project LASY \citep{LASY:2025}, that provides standardization, portability and pre-processing of laser field data for plasma simulation codes, such as FBPIC \cite{Lehe:CPC2016}, HiPACE++ \cite{Diederichs:CPC2022}, WarpX \cite{Fedeli:SC22} and Wake-T \cite{FerranPousa:2019}.

The vacuum propagation routines consist of spectral transforms and the algebraic operations defined \cref{PropEq_vacc3,PropEq_vacc45}. The transforms over time and the Cartesian coordinates are performed via one and two-dimensional fast Fourier transform (FFT) algorithms respectively. Decomposition into the Bessel series for cylindrical geometry, also known as the discrete Hankel transform, consists of the multiplication of radial field distributions $F_m(r_j^{(m)})$ by the transform matrices, $M_{ij}^{(m)}\propto J_m(k_{r, i}^{(m)},r_j)$, where spatial and spectral samplings may depend on the mode number $m$. A standard condition for spectral transforms is the spatial orthogonality of the basis functions (modes) on $r\in [0, R)$, which is imposed by defining $k_{r, i}^{(m)} = \alpha^{(m)}_i / R$, with $\alpha^{(m)}$ being the roots of $m$th order Bessel function. One way to construct the symmetric transform suggested in \cite{guizar:JOCA2004}, involves defining the spatial sampling according to these Bessel roots,  which leads to:
\begin{align}
 M_{ij}^{(m)} = \cfrac{2 \, J_m(\alpha^{(m)}_i \alpha^{(m)}_j / \alpha^{(m)}_{N_r+1})}{\alpha^{(m)}_{N_r+1} \, |J_{m+1}(\alpha^{(m)}_i)| \, |J_{m+1}(\alpha^{(m)}_j)|}\,,\label{guizar_matrix}
\end{align} 
where $N_r$ is a number of the sampling points. \Cref{guizar_matrix} can be applied to both forward and backward transforms using a symmetrization vector $j_i = |J_{m+1}(\alpha^{(m)}_i)|/R$, that serves as either a divisor or a multiplier for the spatial and spectral input fields respectively (see \cite{guizar:JOCA2004} for details).

In multimode calculations, the different spatial samplings of each mode assumed in \cref{guizar_matrix} may be cumbersome as this requires interpolations for the total field reconstruction. As shown in \cite{Lehe:CPC2016}, such transform can also be defined in a more generic way, on a uniform spatial grid. We construct such operator by considering the backward transform defined as,
\begin{align}
 M_{ji}^{(m)(-1)} = J_m(\alpha^{(m)}_i r_j /R)\,,\label{resampling_matrix}
\end{align} 
and obtaining the forward transform $M_{ij}^{(m)}$ via numerical matrix inversion. The sampling $r_j$ is considered uniform and is independent of mode number, and good numerical accuracy is achieved for $R \approx r_{N_r}+\Delta r/2$, where $\Delta r$ is the sampling step.

To solve the plasma propagation equation \cref{PropEq_envelopEq2}, we have explored and implemented a few explicit and implicit numerical schemes. The most suitable methods were found to be the standard 3rd and 4th order explicit Runge-Kutta schemes, and the implicit midpoint and Crank-Nicolson schemes \cite{CrankNicolson:1947}. While the implicit schemes has proven to be more stable, the explicit ones support adaptive step calculations. For these, the new steps are calculated from the empirical formula, ${ \Delta z^{(new)} = \alpha_{err}  \Delta z \sqrt{\varepsilon / \varepsilon_{max}} }$, where the error $\varepsilon$ is the integrated normalized difference between the lower and higher order schemes, and the growth parameter ${\alpha_{err}\lesssim 1}$, can be tuned for stability.

\section{Axiparabola for laser plasma acceleration}\label{sec:sims}

The main objective of the developed numerical tool is the pre-processing and design of advanced laser configurations for simulations in the context of laser plasma acceleration. To demonstrate the use of Axiprop let us consider applications of the axiparabola -- a modified parabolic mirror that produces a quasi-Bessel beam with an extended focal line length and a superluminal flying focus \cite{Smartsev:2019, Oubrerie:JOPT2022}. This optic is attracting a growing interest for LPA, as it can be used either to machine hydrodynamic OFI (HOFI) channels as plasma waveguides \cite{Oubrerie:LSA2022}, or to provide an LPA driver that is immune to diffraction, and can operate without bunch-wake dephasing \cite{Caizergues:NatPhot2020,palastro2020dephasingless}.

\subsection{Axiparabola beam in vacuum}\label{subsec:axibeam}

Let us consider a standard axiparabola, presented by the expanded sag-function \cite{Smartsev:2019}:
\begin{align}
 & s(r) = \cfrac{r^2}{4 f_0} - d_0 \; \cfrac{r^4}{8 f_0^2 R^2} + d_0 \; \cfrac{R^2 + 8 f_0 d_0}{96 f_0^4 R^4} \; r^6\,, \label{axi_sag}
\end{align}
and set the focal distance, the length of the focal depth, and the beam radius to $f_0=500$~mm, $d_0=50$~mm and $R=30$~mm respectively. At the mirror position, the laser has a flat transverse phase, a nearly flat-top super-Gaussian field amplitude profile, ${E\propto\exp(-(r/R)^{30})}$, and a Gaussian temporal profile, ${E\propto\exp(-t^2/\tau^2)}$, with $\tau=27$~fs, and wavelength $\lambda_0 = 800$~nm. The nominal total energy considered in this example is 1~J. 

\begin{figure}[ht]
\centering
\includegraphics[width=0.85\linewidth]{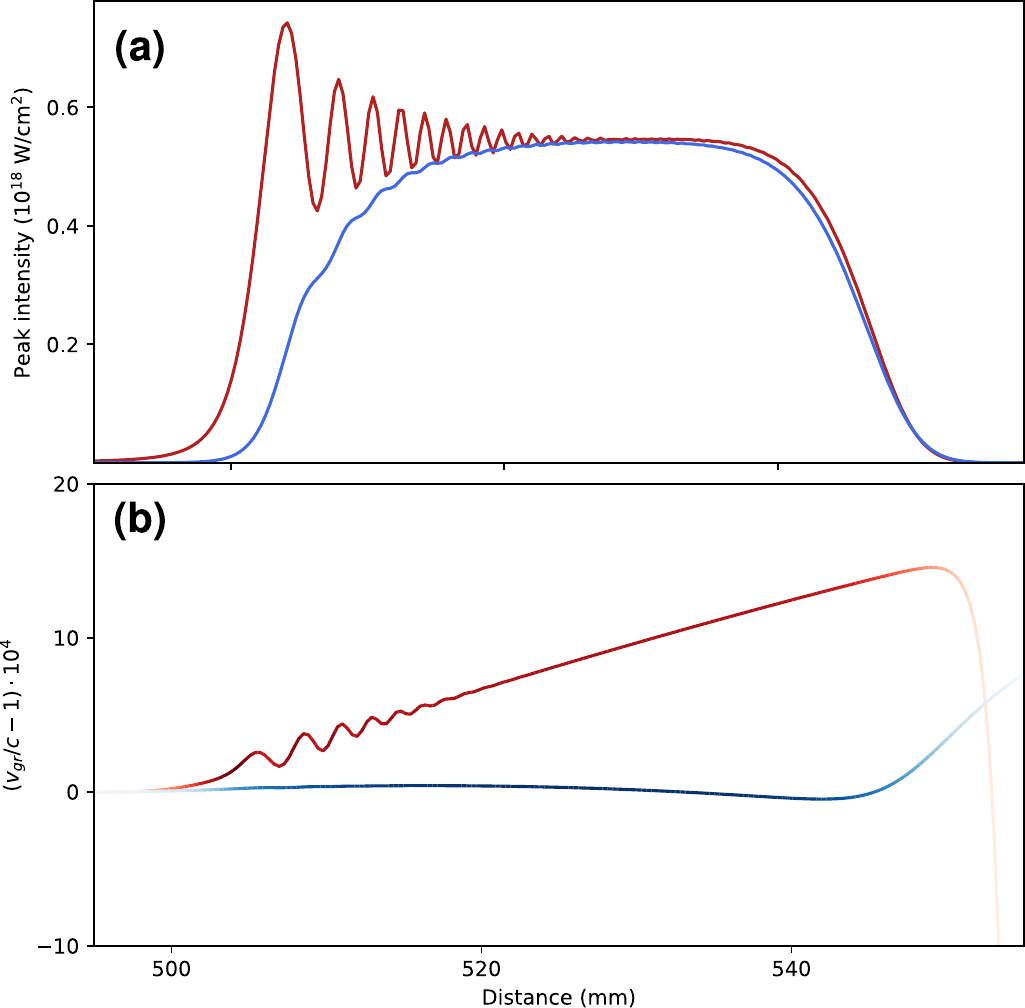}
\caption{Maximum of the on-axis electric field (a), and apparent on-axis pulse group velocity along the focal line (b). Red curves correspond to the standard axiparabola \cref{axi_sag}, blue curve in (a) has the applied attenuation \cref{eq:attenuation}, and blue curve in (b) also has phase correction \cref{pfc_4}. The color intensity in (b) is proportional to the field amplitude.}\label{Fig1}
\end{figure}

The features of interest are the intensity profile and the group velocity of the peak field. In \cref{Fig1}, the red curves show the evolution of peak intensity, and the apparent on-axis pulse group velocity along the focal line. One can see, that the intensity remains high along the distance $d_0$, but exhibits strong oscillations at the beginning and the end of the line. The pulse centre starts with the luminal velocity, but accelerates and quickly becomes super-luminal.

The amplitude modulations at the start of the line result from the interference of the focalized part of the beam, $r\lesssim R$, and the central part of the beam, $r\ll R$, which experiences a nearly parallel propagation, and has a much larger diffraction length. To suppress these modulations the central part of the beam can to attenuated with a Gaussian profile \cite{touguet:arXiv2026},
\begin{align}
E_{att} = E_{0} \left(1-\exp\left(-r^2/R_f^2\right)\right)\,. \label{eq:attenuation}
\end{align}
The blue curve in \cref{Fig1} corresponds to the case of initial beam profile with applied \cref{eq:attenuation} and $R_f=9$~mm, and it shows suppression of intensity modulations along the focal line.

The superluminal driver pulse, presented in red in \cref{Fig1}b, may be impractical for phase-locked LPA, as it produces a few tens of micrometres slippage over the acceleration path, leading to rapid dephasing with the accelerated electron bunch. While plasma itself may compensate for such dephasing by slowing down the pulse, a more explicit control over the pulse acceleration can be achieved through tailored spherochromatism, that can be introduced by an aspheric doublet. For calculations, such coupling can be added in the spectral domain as frequency dependent phase, 
\begin{align}
 & \phi = -\alpha \; (\omega-\omega_0) \; (r/R)^4 \,. \label{pfc_4}
\end{align}

In \cref{Fig1}b, the blue curve shows a clear suppression of the pulse acceleration by applying the phase defined by \cref{pfc_4} with $\alpha=-130$~fs. In the context of LPA, a high-power laser interacts with the plasma and plasma waves, which affects its propagation in a complex way that we will discuss in the following sections. In order to account for such plasma effects and achieve the phase-locked LPA, the parameter $\alpha$ can be adjusted with help of kinetic PIC simulations.

\subsection{Plasma waveguide generation}\label{subsec:channel}

\begin{figure*}[ht]
\centering
\includegraphics[width=0.75\linewidth]{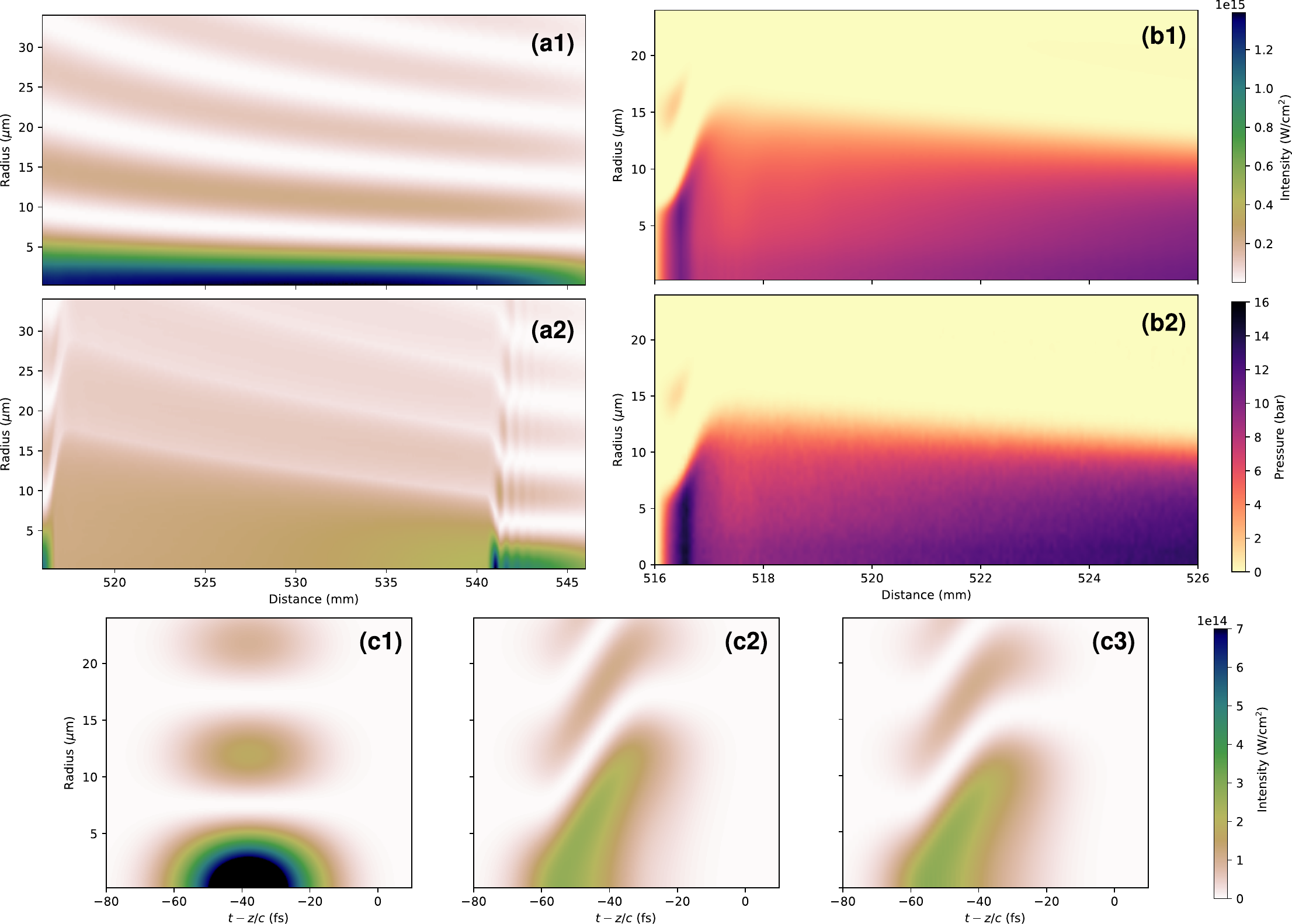}
\caption{(a) Radial profiles of maximum intensity as function of pulse propagation distance in vacuum (a1) and in plasma (a2). (b) Plasma pressure produced by OFI process calculated with Axiprop (b1) and with FBPIC (b2).(b) Intensity spatio-temporal profiles at $z=526$~mm in vacuum (c1), in plasma modelled by Axiprop (c2), and in plasma modelled by FBPIC (c3).}\label{Fig2}
\end{figure*}

Laser guiding is an important method for high power LPA, as it allows to extend acceleration length far beyond the laser diffraction length. One way to create a plasma waveguide is to use a machining laser pulse to ionize a gas column, which then expands to create a density depression with a radius matching the guided beam. The channel characteristics are defined by the radius and electron pressure of the produced plasma, and in OFI process, typical for femtosecond pulses, they depend on the peak laser intensity and the gas species. For the mJ pulses and long multi-cm propagations, OFI may also affect the pulse changing its peak field and intensity temporal and spatial profiles \cite{Monzac:PRR2024}.

The quasi-Bessel beam produced by an axiparabola consists of central spot and surrounding rings, with the intensity of the first ring about $\approx 6$ times lower than the center. For plasma channel generation, the beam should ionize the gas in the central spot, while avoiding ionization at the rings, as the latter would generate the convergent hydrodynamic shocks, and may perturb the resulting channel density profile. The threshold intensity for Coulomb barrier suppression of an ion with ionization potential $U_{ion}$ and ionic charge $Z$ can be estimated as, 
${I\text{[W/cm$^2$]} = 4\cdot 10^{9} \; U_{ion}^4 \text{[eV]}\, / \, Z^2}$ \cite{Augst:PRL1989}.
For hydrogen atoms this gives, ${I_H \simeq 1.4 \cdot 10^{14}}$~W/cm$^2$, and for the configuration considered in Sec.~\ref{subsec:axibeam}, the minimal laser energies required for ionization at the central spot and the first ring are $0.3$~mJ and $1.6$~mJ respectively. In practice, effects such as ionization losses and plasma refraction significantly reduce the on-axis intensity. Consequently, in the simulation we employ a higher laser pulse energy of 2~mJ to maintain sufficient intensity for central ionization. For the propagation medium, we consider atomic hydrogen gas with a number density of ${n_H = 5 \times 10^{18}}$~cm$^{-3}$ situated between $z = 516$~mm and $z = 541$~mm, and featuring a uniform longitudinal profile that starts and ends with 0.5~mm linear ramps.

Note, that the incident rays only interact with the plasma generated by the central spot that has a few micrometres radial size. Therefore, the converging wavefront, extended over a few millimetres diameter,  does not experience plasma diffraction; thus the plasma has no effect on the field group velocity, or the focal line extent. Conversely, as rays cross the plasma near the axis, they experience strong refraction, and the radial size of the central spot increases, while its intensity is decreasing. We demonstrate this effect by comparing the propagation of the laser radial intensity profile in vacuum \cref{Fig2}(a1) and in plasma \cref{Fig2}(a2), and the corresponding spatio-temporal profiles at $z=526$~mm in \cref{Fig2}(c1) and \cref{Fig2}(c2) respectively. From \cref{Fig2}(c1,c2), one may see that, in both cases, the laser pulse front remains unaffected by the plasma until the intensity reaches the ionization threshold $I_H$ at $t-z/c \approx -55$~fs. As the light generates plasma it undergoes strong refraction, and its radial profile significantly expands. This effect is localized within the plasma, and we see in \cref{Fig2}(a1) and (a2), that as the beam exits the plasma its characteristics return to those of vacuum propagation. In practice, this refraction depends on the gas density, laser energy, and axiparabola parameters, thus offering a potential lever for controlling the channel profile.

The carrier-frequency resolved plasma model of Axiprop allows for the calculation of both the electron density $n_e$, and the OFI temperature. The latter is calculated as, $k_b T_e = 2 / 3 \langle W_p \rangle $, where $\langle W_p \rangle$ is the kinetic energy retained by electrons averaged over the cylindrical volume of the numerical cells, ${2\pi r \Delta r}$. Simulations indicate that the hydrogen is fully ionized over the central spot, with electron temperature reaching 2~eV. The resulting plasma pressure $P = n_{e} k_b T_e$ is shown in \cref{Fig2}(b1). To validate the model, this simulation was reproduced using the spectral quasi-cylindrical PIC code FBPIC \cite{Lehe:CPC2016}. The field was generated in Axiprop and converted to the openPMD format via the LASY toolkit \cite{openPMD:Zenodo2015,LASY:2025}. To manage the computational load, the propagation was limited to the range $z=516$~mm to $z=526$~mm. The resulting electrons pressure profile and the laser spatio-temporal intensity profile at $z=526$~mm are shown in \cref{Fig2}(b2) and \cref{Fig2}(c3) respectively. Despite fundamental differences in the underlying models, the two codes demonstrate excellent agreement in both optical and plasma descriptions.

The obtained plasma characteristics can serve for studies of channel formation on nanosecond time scale. At such electron temperatures, plasma expansion into the neutral cold gas is influenced by ionization and recombination, and electron thermal conductivity, which strongly affect the plasma cooling. This process can be modelled using plasma hydrodynamic simulations \cite{Shalloo:PRE2018, Oubrerie:LSA2022, Mewes:PRR2023}. Our previous simulations and experimental measurements \cite{Oubrerie:LSA2022}, have demonstrated that for such parameters, expansion should saturate after 1-2 nanoseconds, yielding a channel with an effective radius of 15-30~$\mu$m, and an on-axis density around $10^{18}$~cm$^{-3}$, matching the requirements for LPA.

\subsection{Flying-focus LPA}\label{subsec:lpa}

Let us consider a femtosecond quasi-Bessel beam used as a diffractionless driver for phase-locked LPA. In order to drive acceleration in the non-linear, blow out regime, the laser has to maintain a relativistically strong amplitude $a_0\simeq 2$ along the focal line. In contrast to the previous $\mu$J laser case, such multi-J pulses ionize the gas over the full aperture, and thus fully experiences ionization, plasma dispersion, and refraction effects, developing temporal and spatial modulations. In particular, around the axis, the laser drives highly non-linear plasma waves with sharp electron density modulations (the so-called bubbles), that refract the converging rays. This latter phenomenon is beyond the scope of the plasma model in Sec.~\ref{subsec:plasma}, which only accounts for ionization and plasma dispersion. While modelling this high-intensity part indeed requires consistent PIC modelling, the main part of converging field is spread over much larger millimetric spots and propagates in the weakly non-linear regime. This suggests that propagation of the pulse front, that does not experience refraction in the wake, should still be consistent with our simple plasma model.

For this study we have performed optical and full PIC simulations with Axiprop and FBPIC respectively. Optical simulations were performed using the enveloped model in \cref{current_motion_rel_env} with AC ADK ionization \cite{ChenM:JCP2013}, and FBPIC simulations were performed using the Lorentz-boost technique with $\gamma_\text{boost}=5$ \cite{Vay:PRL2007,Lehe:PRE2016}. In these simulations we consider a 20~J laser pulse focussed with a holed axiparabola in the same optical configuration as discussed in \cref{subsec:axibeam}. The gas target, consisting of ionizable atomic hydrogen gas with $n_\text{H} = 4 \cdot 10^{18}$~cm$^{-3}$, starts at $z=516$~mm with a 2.5~mm smooth density ramp, followed by a 25~mm constant density plateau, and ends with another 2.5~mm ramp. For this plasma density, we have applied the spatiotemporal couplings in \cref{pfc_4} with  $\alpha=15$~fs in order to produce the plasma wake with nearly constant phase velocity.

\begin{figure}[ht!]
\centering
\includegraphics[width=\linewidth]{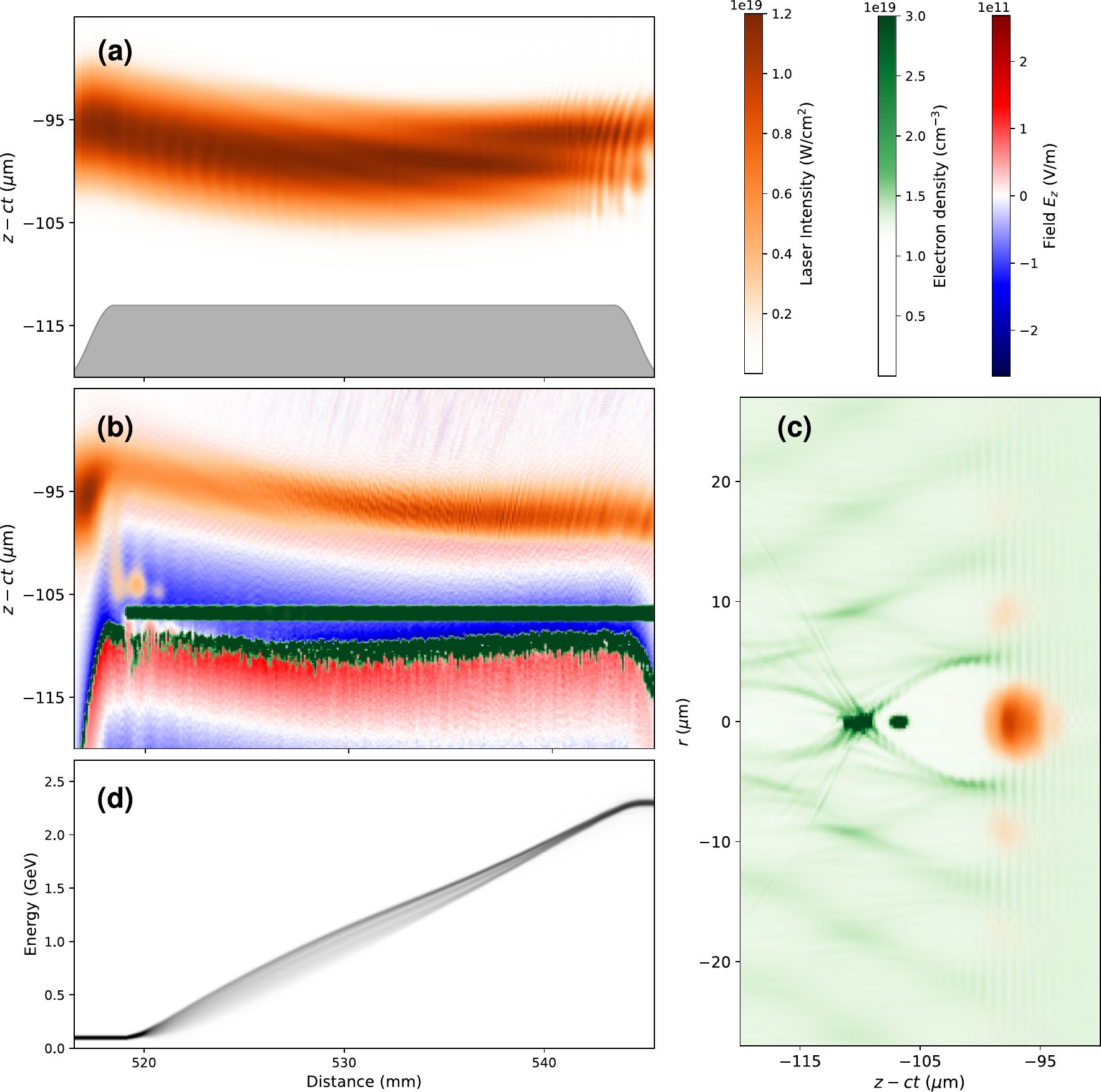}
\caption{(a) Propagation of the on-axis laser intensity profile, ${I(z, \xi=z-ct, r=0)}$, in ionizable plasma with profile shown in grey (scaled) simulated with Axiprop. (b) Propagations of the on-axis laser intensity (orange), electron density (green) and longitudinal field $E_z$ (red-blue) profiles in FBPIC simulation. (c) Instantaneous distributions of laser intensity and electron density after propagation to distance $z=536$~mm. (d) Propagation of the electron bunch spectrum. Colormaps for intensity and electron density values are shared between (a,b,c)}\label{Fig3}
\end{figure}

Propagation of the on-axis intensity ${I(z, \xi=z-ct, r=0)}$ in Axiprop and FBPIC simulations are shown in orange in Figs.~\ref{Fig3}(a) and (b), and the overall profiles show major differences. In the full PIC model (b), as the pulse enters the plasma, the on-axis field gets shortened, while in the optical simulation the pulse duration is preserved. This shortening is explained by the fact, that as the pulse front excites on axis a plasma wave with sharp electron density modulations, it refracts the following rays and prevents them from reaching the axis. In Axiprop, medium motion is not considered, and it describes correctly evolution of the pulse front, but fails at times of plasma wave formation, when a non-linear on-axis current is produced in the motionless plasma. Despite this important discrepancy, the pulse front follows a similar trajectory in both models. In Figs.~\ref{Fig3}(a) and (b) we see the slow temporal modulations that are developed along the propagation, and fast ones arising at the end of it. These modulations result from ionization as confirmed by a comparison with a pre-ionized plasma case in Axiprop, were they are suppressed. The ionization process abruptly changes the dielectric permittivity both temporally and spatially, and its interference with the strong STCs of the considered laser induces the observed complex behaviour. The detailed investigation of this phenomenon is beyond the scope of present work and will be considered in future studies.

In the FBPIC simulation, we visualize the acceleration by injecting a witness electron bunch with sizes $\sigma_z = 0.4$~$\mu$m, $\sigma_r = 0.25$~$\mu$m, an energy 100~MeV with 0.1\% spread, and a normalized emittance that matches the focussing plasma fields, ${\epsilon_n=k_p\sigma_r^2\sqrt{\gamma/2}}$. In \cref{Fig3}(b), we plot the on-axis electron density (green) and the longitudinal field distributions (red-blue). The electron density distribution comprises the witness beam which propagates at a nearly constant luminal velocity (straight green stripe at $\xi=-106$~$\mu$m), and a second peak corresponding to the far end of the ion cavity, where electrons forming the wave fall on axis. In \cref{Fig3}(c), we plot the corresponding instantaneous spatial distributions of the laser field and electron density at $z=536$~mm, the discussed on-axis features can be identified. A plasma wave driven by a Bessel beam in \cref{Fig3}(c) demonstrates the typical complex multi-ring, similar to the one observed experimentally \cite{Liberman:NatComms2025}.

In \cref{Fig3}(b) we see, that while the witness bunch globally remains in the accelerating field (blue color), the phase velocity of the plasma wave exhibits minor drifts. This dynamics results from the combination of effects, such as driver spot-size dynamics, non-perfect compensation of the optically produced STCs and plasma dispersion, ionization-induced front modulations. In strong, micrometric plasma waves, the longitudinal electric field is distributed linearly on around the bubble center, but becomes highly non-linear and fast growing in the vicinity of the electron spike at the back of the bubble. For electrons travelling in this field, the phase modulations have significant effect on acceleration quality, and to demonstrate it we plot the evolution of the bunch spectrum along the propagation in \cref{Fig3}(d). As we see from \cref{Fig3}(b), when the bunch enters the plasma wake it first appears very close to the bubble boundary, where the non-linear field quickly drives higher energies to the bunch front, thus increasing the total energy spread observed in \cref{Fig3}(d). As LPA proceeds, the phase of the wake shifts and electrons propagate in the linear field that gradually counteracts this chirp, and the spectrum is partially recompressed down to 2.7\% spread (rms) as its mean energy reaches 2.3~GeV.

Our study indicates, that flying-focus LPA is rich in complex optical and plasma phenomena that determine the accelerator performance. The presented configuration is developed with help of multiple Axiprop and FBPIC runs for the demonstrative purpose, but is not specifically optimised for efficiency. Fine-tuning can be done with advanced spatio-temporal shaping as discussed in \cite{touguet:arXiv2026,touguet:arXiv2026b}, and will be the subject of the further investigations. 


\section{Summary}

In this paper we have discussed methods for optical modelling of ultra-short laser pulses with complex spatio-temporal shapes, and their propagation in weakly relativistic ionizable plasmas. These methods were implemented for numerical simulation in an open-source simulation library Axiprop, available to the community and open for contributions. The tool is developed in the context of laser plasma acceleration, and relevant studies of laser guiding, and we have presented and discussed examples of numerical studies, that implement the axiparabola mirror for plasma waveguide generation, and for phase-locked laser plasma acceleration. 

\section{Acknowledgements}

We are grateful to Jerome Faure (LOA), Slava Smartsev (IJCLab), Anton Golovanov (WIS), and to all members of LASY team \cite{LASY:2025}, and in particular Maxence Thevenet (DESY), Remi Lehe (LBNL), Rob Shalloo (DESY) for useful discussions and contributions. We also acknowledge the Grand Équipement National de Calcul Intensif (GENCI) and Très Grand Centre de Calcul (TGCC) for granting us access to the supercomputer Joliot-Curie under the Grant No. 2023-A0150510062 to run the PIC simulations.

\bibliographystyle{unsrt}
\bibliography{biblio}

\end{document}